\documentclass[letterpaper,10pt]{article}
\pdfoutput=1

\usepackage{amsmath}
\usepackage{amssymb}
\usepackage{latexsym}

\usepackage[pdftex]{graphicx}

\usepackage{cite}

\usepackage{color}
\usepackage{multirow} 
\usepackage[table]{xcolor}
\usepackage{rotating}
\usepackage{fancyhdr}
\usepackage{authblk}
\usepackage{url}

\usepackage{booktabs}

\usepackage[labelfont=bf,labelsep=period,justification=raggedright]{caption}

\makeatletter
\renewcommand{\@biblabel}[1]{\quad#1.}
\makeatother

\title{\vspace{-2cm} Correlations between user voting data, budget, and box office for films in the Internet Movie Database}
\author[1]{Max Wasserman}
\author[2,3]{Satyam Mukherjee}
\author[2]{Konner Scott}
\author[2]{Xiao Han T. Zeng}
\author[2,4,$\dagger$]{Filippo Radicchi}
\author[2,3,4,5,*]{Lu\'{\i}s A. N. Amaral}
\affil[1]{\normalsize{Department of Engineering Sciences and Applied Mathematics, Northwestern University}}
\affil[2]{Department of Chemical and Biological Engineering, Northwestern University}
\affil[3]{Northwestern Institute on Complex Systems (NICO), Northwestern University}
\affil[4]{Howard Hughes Medical Institute (HHMI), Northwestern University}
\affil[5]{Department of Physics and Astronomy, Northwestern University, Evanston, Illinois}
\affil[$\dagger$]{Present address: Center for Complex Networks and Systems Research, School of Informatics and Computing, Indiana University, Bloomington, Indiana}
\affil[*]{To whom correspondence should be addressed; E-mail:  amaral@northwestern.edu}
\date{}

\begin{document}

\maketitle

\section*{Abstract}

The Internet Movie Database (IMDb) is one of the most-visited websites in the world and the premier source for information on 
films. Like Wikipedia, much of IMDb's information is user contributed. IMDb also allows users to voice their opinion on the 
quality of films through voting. We investigate whether there is a connection between this user voting data and certain economic 
film characteristics. To this end, we perform distribution and correlation analysis on a set of films chosen to mitigate effects 
of bias due to the language and country of origin of films. We show that production budget, box office gross, and total number of 
user votes for films are consistent with double-log normal distributions for certain time periods. Both total gross and user votes 
are consistent with a double-log normal distribution from the late 1980s onward, while for budget, it extends from 1935 to 1979. In 
addition, we find a strong correlation between number of user votes and the economic statistics, particularly budget. Remarkably, 
we find no evidence for a correlation between number of votes and average user rating. As previous studies have found a strong 
correlation between production budget and marketing expenses, our results suggest that total user votes is an indicator of a film's 
prominence or notability, which can be quantified by its promotional costs.


\section*{Introduction} 

In today's world, we are seemingly in constant connection to the Internet. Most of our activities are stored in electronic databases, 
and the aggregation of all this information represents a novel source for the study of human behavior~\cite{Castellano2009}. 
Indeed, researchers have reported on the statistical properties of the communication patterns of 
e-mail~\cite{Ebel2002,Malmgren2008,Radicchi2009} and traditional `snail' mail~\cite{Malmgren2009}, on the analysis of the 
macroscopic features of web surfing~\cite{Johansen2001,Goncalves2008}, and so on. Aggregate electronic information has not 
only been a boon to scientific investigation, it has also demonstrated utility in many practical applications. For instance, 
aggregate information is used by eBay (\texttt{ebay.com}) to quantify the reputation of sellers and buyers, and researchers have 
used data from Twitter (\texttt{twitter.com}) to analyze collective moods~\cite{Golder2011} and monitor the spread of 
ideas~\cite{Aral2009}.

The information present in the Web is the result of the aggregation of the work of many individuals as well as the outcome 
of complex and self-organized interactions between large numbers of agents. For example, the vast amount of information contained 
on Wikipedia (\texttt{wikipedia.org}) is the product of millions of user contributions. The collaborative and collective outcome is 
not merely the sum of the knowledge of each individual contributor, it is also the result of continuous modifications and refinements 
by users. The content generated through this collaborative strategy is generally more complete than those produced by individuals 
because the collaborative framework ensures more control of the quality of the provided information~\cite{Wuchty2007,Woolley2010}.

The Internet Movie Database (IMDb, \texttt{www.imdb.com}), one of the most frequently-accessed websites worldwide~\cite{AlexaImdb}, 
is home to the largest digital collection of metadata on films, television programs, videos, and video games. Similar to Wikipedia, 
IMDb's content is updated exclusively by unpaid registered users. In addition to accepting user-contributed information, IMDb also 
allows users to rate on a 1 to 10 scale the quality of any film or program. Adding new information to the database is a mostly 
altruistic activity, as it requires action on the part of the contributor in order to enhance the understanding of others. However, 
voting on the quality of a film is a less altruistic action as the user is able to voice his or her opinion through voting. Is 
there any useful information to be drawn from online voting information? IMDb is not a new subject for scientific analysis. It 
has been used in the context of studying the actor collaboration network~\cite{Watts1998,Amaral2000,Herr2007} and in developing 
recommendation systems~\cite{Grujic2008}. More recently, IMDb's extensive collection of keywords was used to create a metric of film 
novelty~\cite{Sreenivasan2013}. However, little work has been done on the connections between the user contributions to IMDb's 
database of information and user contributions to a film's rating score.

In this paper, we seek to identify correlations between IMDb's user ratings of films and various other characteristics reported in 
the database. To accomplish this, we use information available on IMDb to construct a directed network of films. This network
provides the dataset for our analysis. We proceed to filter the network to account for biases on the part of users. Using 
metadata collected on our filtered dataset, we determine the distributions of various characteristics in several time windows in 
order to identify temporal changes. In addition, we perform linear regressions with the goal of quantifying correlations between 
user voting data and user-contributed information.

\section*{Data}

We limit our analysis of IMDb data to films, including in our study both feature-length and short films. We choose to include 
short films as excluding them would ignore almost all films made before 1920. We retrieved data from IMDb on October 26th, 2012, 
and therefore, only include in our analysis films released by 2011. Note that we do not exclude documentary films.

\begin{figure}[!ht]
\centering
\includegraphics[width=0.96\textwidth]{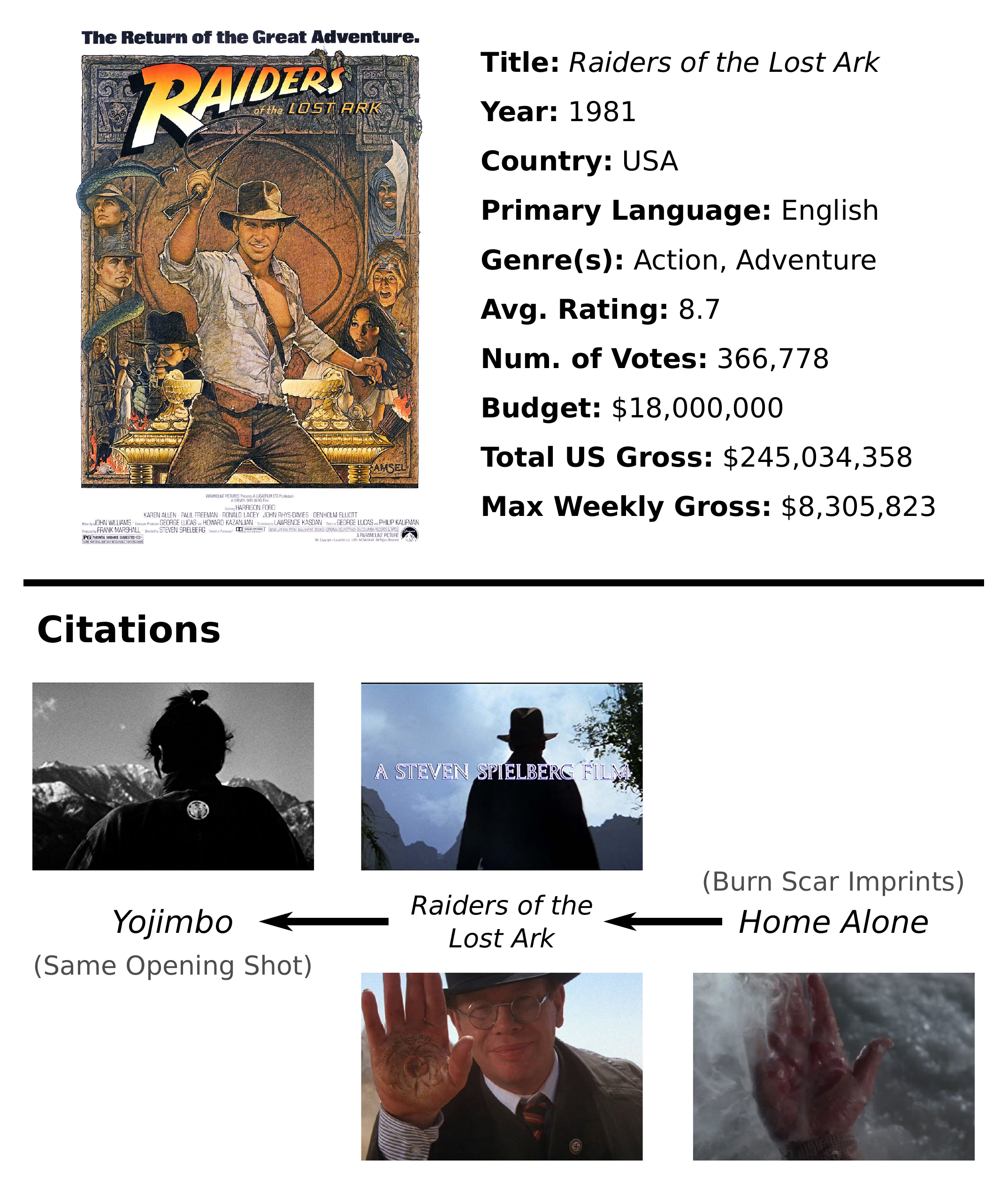}
\caption{
\textbf{Typical metadata on films included in IMDb.}
\textbf{(A)} Examples of metadata collected for \textit{Raiders of the Lost Ark}.
\textbf{(B)} A depiction of two directed edges in the connections network. One edge represents a citation by 
\textit{Raiders of the Lost Ark} to \textit{Yojimbo} (the opening shot of the former honors the opening shot of 
the latter). The other edge represents a citation linking \textit{Home Alone} to \textit{Raiders} (villains in both 
films suffer burns to the hand that leave an impression).
}
\label{fig-example}
\end{figure}

The metadata on films included in IMDb consists of year of release, country of production, primary language, user voting 
statistics, and several types of financial information (Fig.~\ref{fig-example}). All metadata is user-edited apart from the voting 
data, for which IMDb automatically tallies the total number of user votes and reports an average rating using an ``in-house 
formula''. From among all types of financial information reported for a film, we focus on production budget, box office gross in 
the United States, and greatest amount grossed in a single week during a film's theatrical run. All of these values are in unadjusted 
US dollars, and thus not corrected for inflation or GDP growth.

Among the plethora of available and editable information is a section titled ``connections,'' a list of references and 
links between films and other media. All connections listed on IMDb are classified as one of eight ``types'': References, spoofs, 
features, follows, spin-offs, remakes, versions, and edits. We only consider the connections that are classified as references, 
spoofs, or features. A \emph{reference} is a connection between films where one contains an homage to the other in some form. 
For example, the famous flying bicycle scene in \textit{E.T.: The Extra-Terrestrial} is a reference to a sequence in \textit{The 
Thief of Bagdad} where characters also fly in front of the moon. References also come in the form of similar quotes, similar 
settings, or similar filming techniques. A \emph{spoof} is a connection between films where one mocks the other. For example, the 
wagon circle scene from \textit{Blazing Saddles} is a spoof of the final scene from \textit{Stagecoach}. A \emph{feature} occurs 
when a film includes an extract from another film in it. The scene in \textit{When Harry Met Sally} where the title characters watch 
\textit{Casablanca} is an example of a feature connection. Our analysis is limited to these connections as they pertain to parts of 
films, such as scenes or quotes, and thus are conscious choices on the part of creative people such as directors, actors, and 
screenwriters.

\begin{figure}[!ht]
\centering
\includegraphics[width=0.85\textwidth]{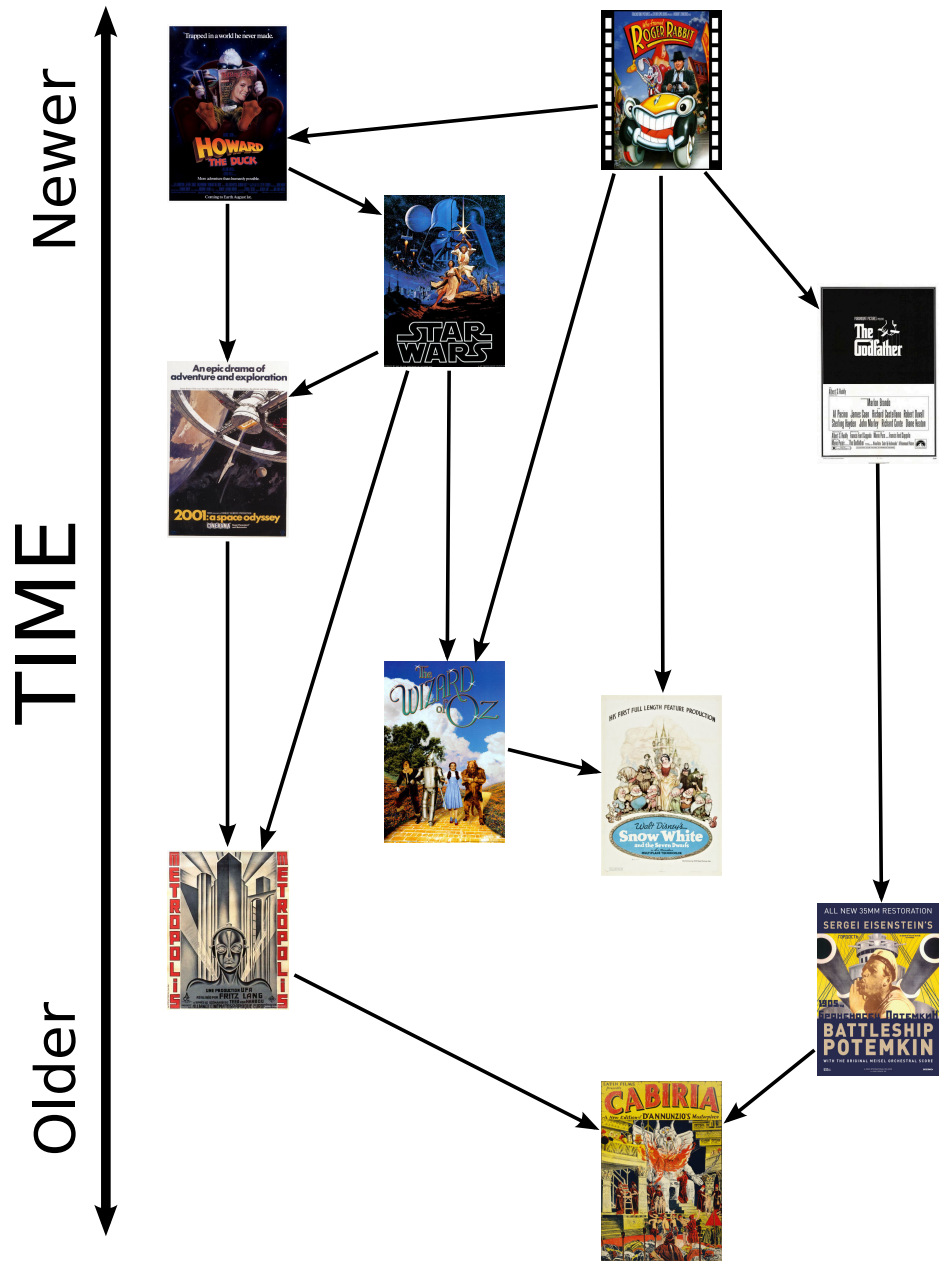}
\caption{
\textbf{Sub-graph of film connections network.}
A sub-graph containing 10 films out of the 28,743 in the giant component of the film connections network. Films are ordered 
chronologically, based on year of release, from bottom to top. A connection between two films means that a sequence, sentence, 
character, or other part of the referenced film has been adopted, utilized, or imitated in the referencing film. For example, 
there is a connection from \textit{Star Wars} (third from the top) to \textit{Metropolis} (third from the bottom) because 
C-3PO is modeled on the robot from Fritz Lang's 1927 film. 
}
\label{fig-network}
\end{figure}

Using the connections between films, we construct a network where each film is a node and each connection is an arc 
(Fig.~\ref{fig-network}). An arc connecting movie A to movie B indicates that movie A contains a reference to movie B (or spoofs 
movie B, or features a clip from movie B). We admit a connection into the network only if the citing film was released in a later 
calendar year than the cited film; that is, all of the links in our network are directed backward in time and the network contains 
no links between two films released in the same calendar year. This constraint ensures that the network is acyclic. The network 
we construct from the metadata on film connections consists of 32,636 films and 77,193 connections.

We limit our analysis to the largest weakly-connected component of the network formed by film connections. This \emph{giant 
component} consists of 28,743 films (88\% of all films in the network) linked by 74,164 arcs (96\% of all connections). For each 
film, we take note of its number of incoming and outgoing arcs. In network theory, these values are known, respectively, as the 
\emph{in-degree} and \emph{out-degree} of a node. It is desirable that ``connections'' is a relatively recondite category in IMDb, 
as the presence of reported connections functions as a minimum threshold for consideration in our analysis. In addition, we 
assume that any film with a nonempty connections section has sufficient information in sections that are better known, such as box 
office information.

\subsection*{Biases in metadata reporting}

Because IMDb is user-edited, we must investigate possible biases in reporting. While user editing allows a reference website such 
as IMDb to be up-to-date, it diffuses the responsibility for fact checking, leading to greater uncertainty about accuracy and 
objectivity of information. Biases may be due to the make-up of the user base. For example, Wikipedia recently reported that 91\% 
of its user editors are male~\cite{WMF2011}, which probably explains why female-focused topics are less thoroughly covered. Therefore, 
we evaluate the basic properties of a database in order to account for biases prior to performing a comprehensive analysis.

Two characteristics of a film that could reveal biases of the user editors are \emph{country of production} and \emph{primary 
language}. Thus, we assign films in the giant component to one of three groups based on country and language. Films produced 
in the United States, regardless of language, are assigned to the ``USA'' group. Films made outside the USA with English as the 
primary language are assigned to the ``English Non-USA'' group. Films produced outside the USA and in a language other than 
English are designated as the ``Non-English Non-USA'' group. Note that the USA group comprises a majority of the films in the 
connections network.

\begin{figure}[!ht]
\centering
\includegraphics[width=0.87\textwidth]{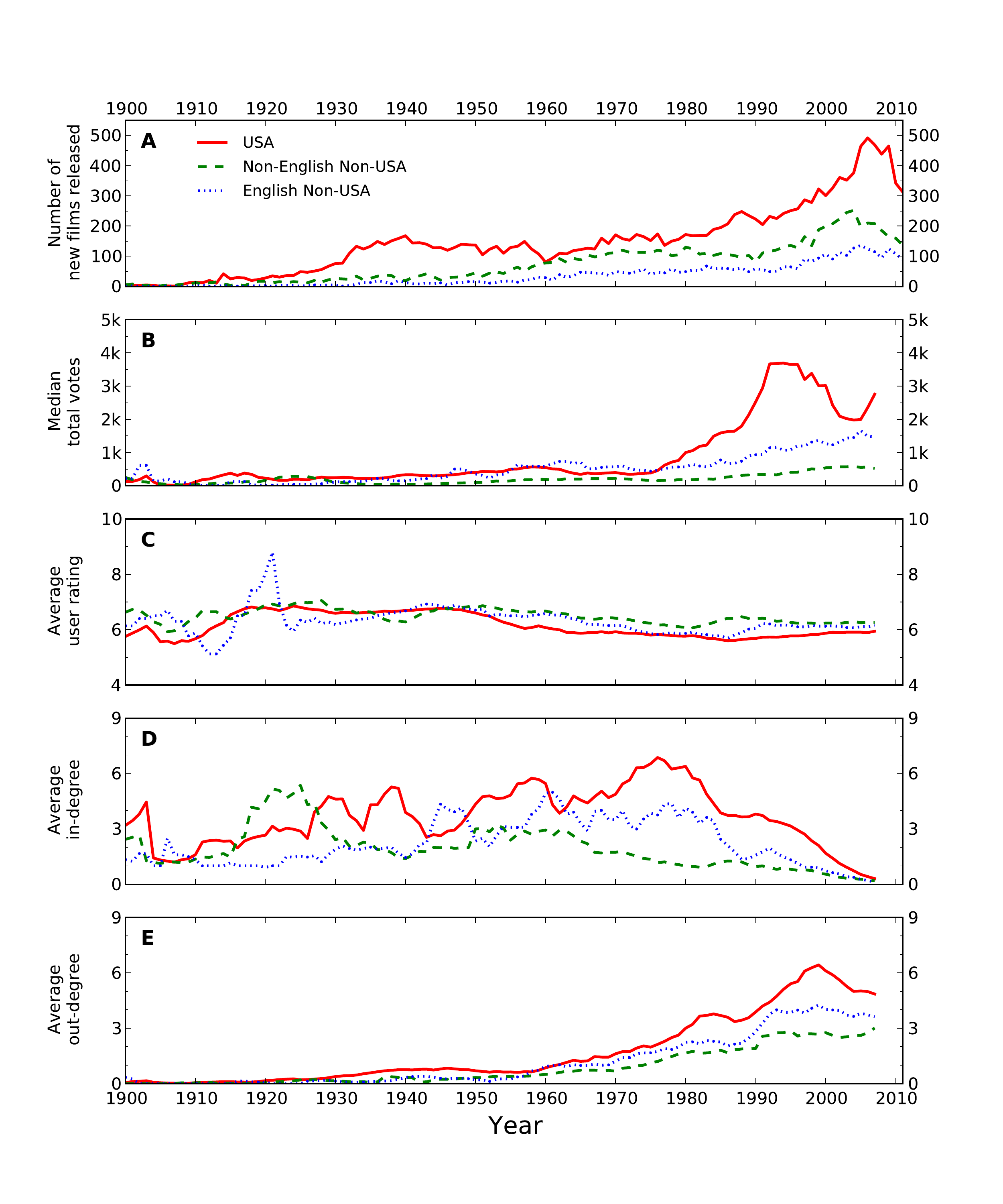}
\caption{
\textbf{Characteristics of movies in the giant component.}
We partitioned films in the giant component of the connections network into three groups: 
USA films, English Non-USA films, and Non-English Non-USA films.
\textbf{(A)} Time dependence of number of films released annually.
Time dependence over five-year windows of
\textbf{(B)} median number of votes, 
\textbf{(C)} average user ratings,
\textbf{(D)} average in-degree, and
\textbf{(E)} average out-degree.
We used five-year windows to calculate all statistics apart from number of films released because of data variability 
on a year-to-year basis. We show the median for \textbf{(B)} because the relatively few films with large numbers 
votes (i.e., 100,000 or more) skew the mean heavily, making it less representative of a typical film.
}
\label{fig-temporal}
\end{figure}

In Fig.~\ref{fig-temporal}, we consider the time dependence of various properties for each group. The number of new films released 
annually increases over time for each of the three groups, particularly rising during the last two decades (Fig.~\ref{fig-temporal}A). 
The decrease in new films from 2009 to 2011 is presumably caused by the recession of 2008--2010, but it may also be due to a 
reporting delay for low budget films.

There is a stark difference in the number of votes films receive depending on their year of release and grouping 
(Fig.~\ref{fig-temporal}B). USA films released after 1990 have a median of more than 2,000 user votes, while those released before 
1980 have a median below 500. Additionally, there is a trend-reversing dip in the median number of votes received by films beginning 
around 1995. We attribute this reversal to a sizable jump between 2003 and 2008 in the number of new films released, which we presume 
are mostly independently-produced, ``obscure'' pictures that likely receive few votes on IMDb. 

Films in the English Non-USA group receive more votes than films from the Non-English Non-USA group, despite Non-English Non-USA 
films outnumbering English Non-USA films by almost 2 to 1 (Fig.~\ref{fig-temporal}B). Similarly, we observe that the English Non-USA 
group averages more incoming and outgoing citations than the Non-English Non-USA group despite the latter being more numerous in the 
dataset (Fig.~\ref{fig-temporal}D-E). These findings suggest there is both a language bias and a temporal bias in the distribution of 
user votes in IMDb. These biases are not observed in the average user ratings for films, as there is little change over time and 
among the three groups (Fig.~\ref{fig-temporal}C).

Surprisingly, the average in-degree declines for films released after 1992, while the average out-degree increases for films 
released after the mid-1980s (Fig.~\ref{fig-temporal}D-E). The latter is to be expected because connections between films can only 
travel ``backwards'' in time, i.e., from a newer film to an older film. Unexpectedly, the average out-degree declines for films 
made after 1999, particularly for films in the USA group.

The presence of a temporal bias towards recent films in the IMDb connections network is not unexpected, as modern technology 
allows films to be produced more quickly and potentially less expensively than ever before. This trend in costs makes the 
downward trend in average out-degree for films beginning around 1999 stand out. It is unlikely that films released in 1998 are 
more citation-laden than films released in 2005. Instead, we presume that this decrease indicates a different sort of temporal 
bias, wherein IMDb does not yet have full information for films made in the 2000s while information for films made in the 1990s 
is more complete. Thus, the data suggest that there is a latency period for the reporting of film connections, a fact that must 
be taken into consideration in interpreting any analysis.

In order to better quantify biases in enumeration of connections, we consider the observed proportions of incoming and outgoing 
connections for each group. We perform a series of Monte Carlo simulations wherein films are randomly assigned to one of the three 
groups while each total group size remains constant. In this way, we are able to calculate the expected proportions of incoming 
and outgoing connections of each group if countries and languages of films were unimportant. If the randomized proportions are 
significantly different from the actual proportions, we must conclude that there are country and language biases.

\begin{figure}[t]
\centering
\includegraphics[width=1.0\textwidth]{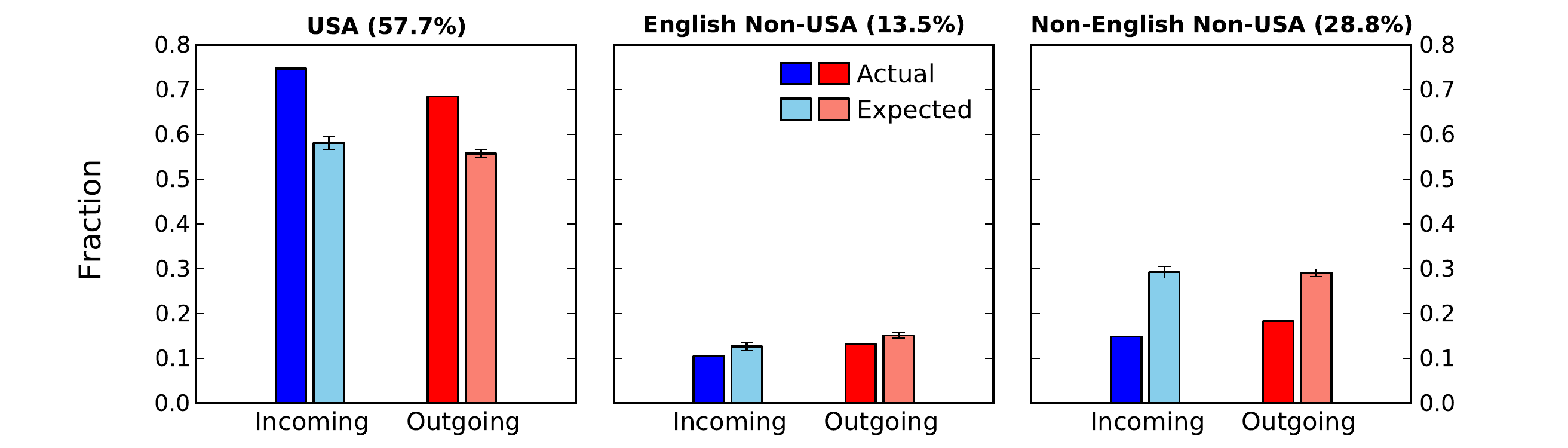}
\caption{
\textbf{Actual and expected fractions of connections by country/language grouping.}
Fractions of incoming connections and outgoing connections in the giant component of the film connections network for USA 
films (left plot), English Non-USA films (center), and Non-English Non-USA films (right). The numbers in parentheses 
represent the percentages of nodes in the network that belong to each group. Dark blue and dark red bars represent the 
fractions of connections in the giant component. Light blue and light red bars represent the average fraction of 
connections in the giant component following a Monte Carlo simulation where the films were randomly reassigned to one of 
the three country/language groupings. Error bars represent one standard deviation of the mean following 10,000 simulations.
}
\label{fig-countrybars}
\end{figure}

\begin{table}[ht]
\centering
\begin{tabular}{lcccc}
& & \multicolumn{3}{c}{\% of links from} \\
\cmidrule(lr){3-5} 
& & & \makebox[1.3cm][c]{English} & \makebox[1.3cm][c]{Non-Eng.} \\
Group & \% of nodes & \makebox[1.3cm][c]{USA} & \makebox[1.3cm][c]{Non-USA} & \makebox[1.3cm][c]{Non-USA} \\
\midrule
USA & 57.7 & 84.5 & 64.8 & 45.1 \\
English Non-USA & 13.5 & 8.8 & 21.8 & 8.4 \\
Non-English Non-USA & 28.8 & 6.7 & 13.4 & 46.5 \\
\midrule
Total & 100.0 & 100.0 & 100.0 & 100.0 \\
\end{tabular}
\caption{
\textbf{Bias due to country of production and primary language.}
Note how USA films mostly reference other USA films (84.5\% of connections point to 57.7\% of the nodes) while mostly ignoring 
Non-English Non-USA films (6.7\% of the connections point to 28.8\% of the nodes).
}
\label{tbl-connfrac}
\end{table}

Our analysis reveals that USA films have a disproportionate fraction of incoming and outgoing connections (Fig.~\ref{fig-countrybars}). 
This is a strong indicator of the USA and English-language biases present in the dataset. The USA bias is further evident when we perform 
second-order analysis on the fractions of connections. We find that USA films receive percentages of the outgoing connections from other 
USA films and English Non-USA films that are greater than the percentage of USA nodes in the network (Tbl.~\ref{tbl-connfrac}). 
Additionally, Non-English Non-USA films cite USA films nearly as often as they cite other Non-English films (Tbl.~\ref{tbl-connfrac}). 
Although this is clear evidence of bias for American films, it is not necessarily indicative of an American bias in IMDb's user base. 
IMDb is not an American-centric website as it was originally founded in the United Kingdom~\cite{Needham2010}. Moreover, the United 
States only accounts for 31\% of IMDb's total traffic~\cite{AlexaImdb}. (For comparison, the USA comprises 30\% of Google's traffic and 
30\% of Apple's traffic~\cite{AlexaGoogle,AlexaApple}.) More likely, the over-representation of USA films in the network reflects the 
pervasiveness of the American film industry around the world. For example, between 2007 and 2009, 23 of the 27 most-viewed films 
worldwide originated from the United States and the other 4 were co-produced by United States companies~\cite{Acland2012}. As such, 
more users are likely to identify citations to and from American films because they are the most-viewed films worldwide.

\begin{table}[tb]
\centering
\begin{tabular}{rccc}
& & & Giant Component \\
& Entire Network & Giant Component & of USA Group \\
\midrule
Number of films & 32,636 & 28,743 (88\%) & 15,425 (47\%) \\
Number of connections & 77,193 & 74,164 (96\%) & 42,794 (55\%) \\
\end{tabular}
\caption{
\textbf{Characteristics of film connection networks.}
Number of films and connections in the entire network, the giant component of the entire network, and the giant component of the network 
of films in the USA group only. Numbers in parentheses indicate the percentage of nodes or edges in the entire network. 
}
\label{tbl-networks}
\end{table}

Due to the American-centric nature of both the financial information and the connections network as a whole, we choose to restrict 
our focus to films produced in the USA. This choice will remove confounders caused by country of production from our analysis. 
Our new dataset is the giant component of the connections network for USA films, which consists of 15,425 films and 42,794 
connections (Tbl.~\ref{tbl-networks}).

\subsection*{Missing information}

\begin{figure}[t]
\centering
\includegraphics[width=0.6\textwidth]{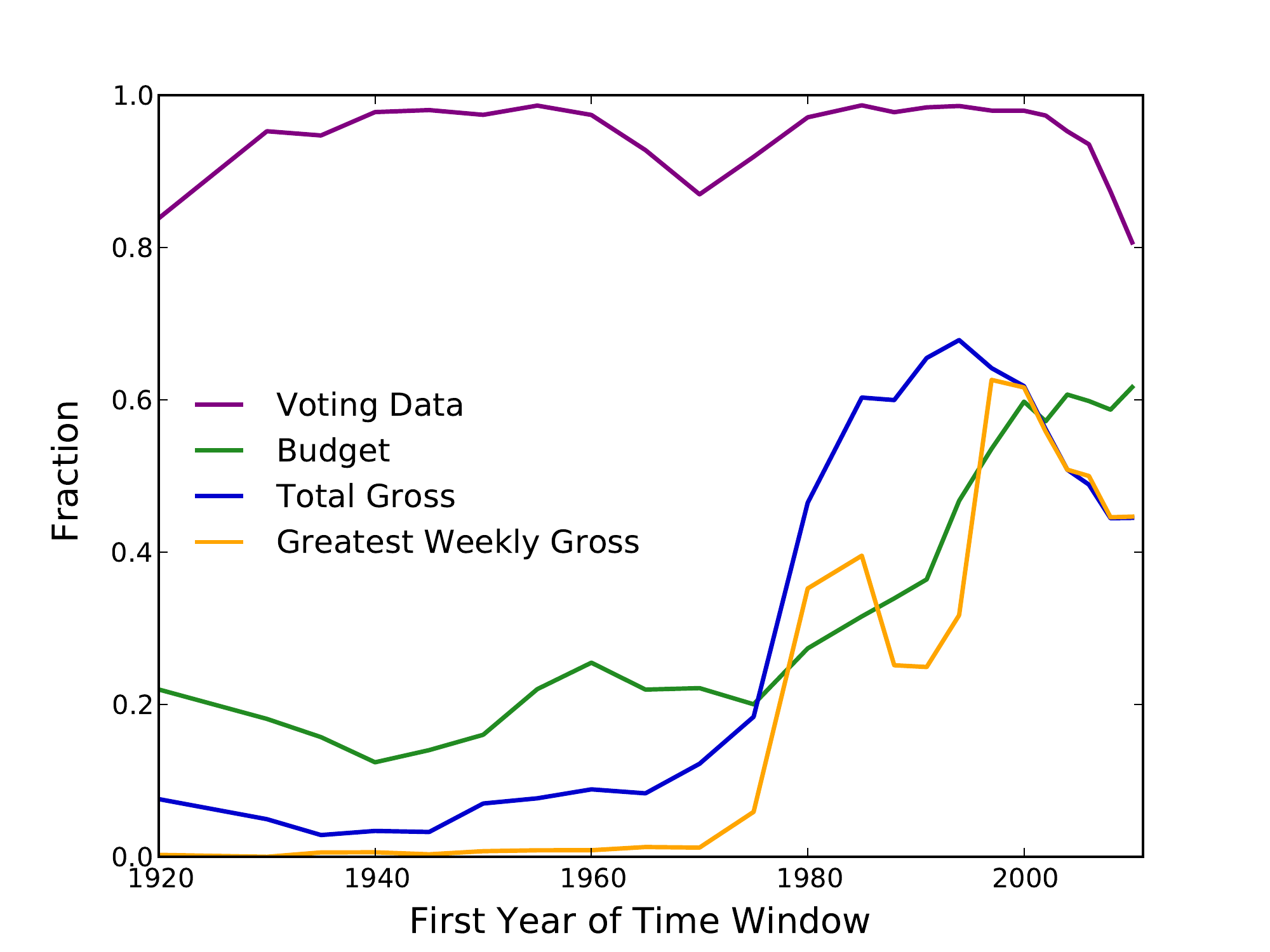}
\caption{
\textbf{Prevalence of reported data for USA films.}
}
\label{fig-datafrac}
\end{figure}

Not all IMDb entries report budget, total gross, or weekly gross information. Missing data increases for older films, with 
``greatest weekly gross'' being the most affected category of data (Fig.~\ref{fig-datafrac}). It appears that the weekly box office 
take may not have been a regularly reported statistic until the 1980s, as that is the time when a sizable increase occurs in the 
reporting of greatest weekly gross. In fact, the weekly gross data found on Box Office Mojo---a website affiliated with IMDb---cover 
only the period after 1980~\cite{BoxOfficeMojo}. Due to the lack of reported weekly gross data prior to 1975, we choose to omit 
greatest weekly gross from our subsequent analysis.

Some IMDb entries also do not report a film's average user rating or number of user votes received. This reflects part of IMDb's 
methodology for calculating average user rating as the website does not post voting figures until a film has received a minimum 
of five user votes. Recent films are the most likely to lack voting data (Fig.~\ref{fig-datafrac}), which may be explained by 
users waiting to rate films until after viewing them.

\section*{Results}

We proceed to examine the distributions of the values for three film statistics: Production budget, total gross, and total number of 
user votes. We study the logarithm of these quantities because their values span several orders of magnitude. In order to minimize 
the effects of temporal bias, we look at the distributions for sets of films within 23 time windows spanning the period 1920--2011. 
We vary the length of the time windows in order to ensure that the number of films in each window is approximately constant. In 
addition, looking at financial values within narrow time windows mitigates the effects of GDP growth.

For each time window, we find the best-fitting Gaussian and double-Gaussian distribution parameters for the relevant data values. 
We then use bootstrapping to determine the statistical significance of the fits.

Because the double-Gaussian model is defined as a linear combination of two ``single'' Gaussian models---thus having five parameters 
instead of two---the double-Gaussian model will provide a better fit for the data under most circumstances. Therefore, in order to 
accept the double-Gaussian model as the true distribution, we must reject the Gaussian model as a possible fit and fail to reject 
the double-Gaussian model. If we fail to reject the Gaussian model, we assume that it is the correct description for the data. Using 
the Bonferroni correction, the threshold for rejecting a model is $p_{\text{B}} = 0.00217$~\cite{Shaffer1995}.

\begin{figure}[tbp]
\centering
\includegraphics[width=1.0\textwidth]{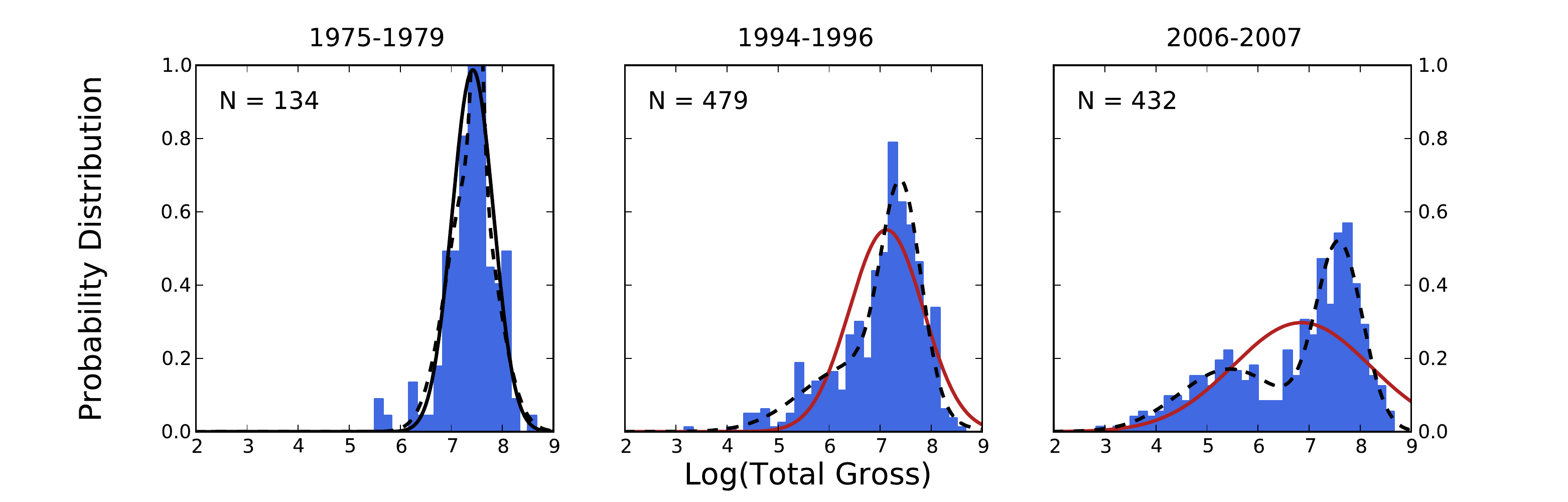}
\caption{
\textbf{Examples of distributions of log total gross.}
Distribution of the log of total US gross in different time windows. The lines represent the best fits of a single 
Gaussian distribution (solid line) and a mixture of two Gaussian distributions (dashed line) to the data. The color of the 
curve signifies whether we reject (red) or fail to reject (black) the distribution as a possible fit for the data. The number 
in the upper-left corner is the total number of data points in the sample.
}
\label{fig-hist-gross}
\end{figure}

\begin{figure}[tbp]
\centering
\includegraphics[width=1.0\textwidth]{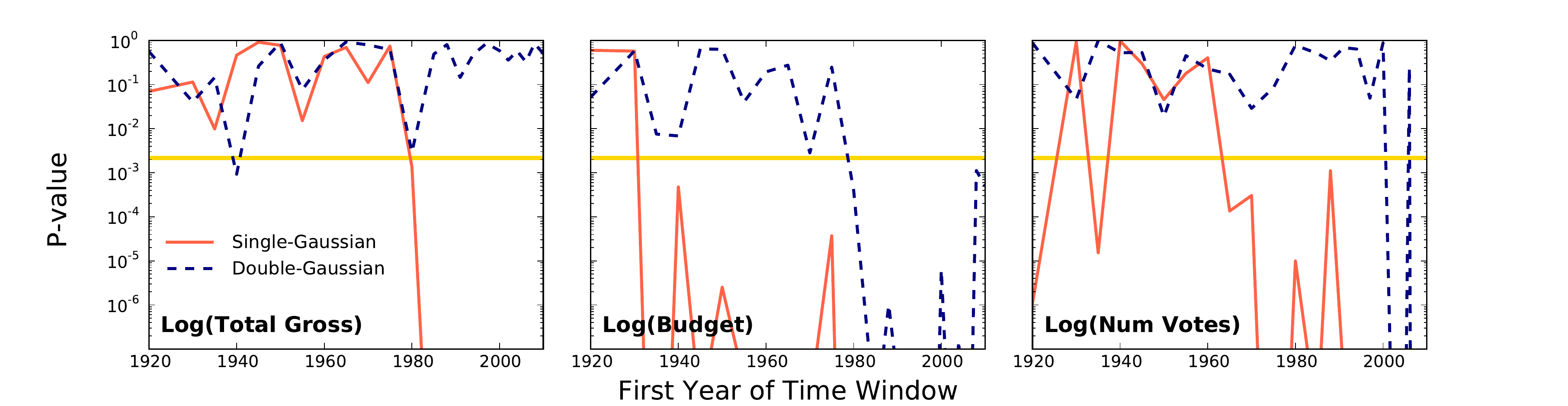}
\caption{
\textbf{P-values of distribution fits over time.}
P-values from bootstrapping analysis representing the goodness of fit of single-Gaussian and double-Gaussian distributions in 
specific time windows for the four considered statistics: The log of total gross (left), the log of film budget (center), and 
the log of number of user votes (right). The gold line represents the threshold P-value---calculated according to the Bonferroni 
correction---below which we reject the distribution. 
}
\label{fig-bootstrap}
\end{figure}

Among the statistics, the logarithm of total US gross has the strongest evidence for the double-Gaussian distribution 
(Fig.~\ref{fig-hist-gross}). We take the Gaussian model as its best representation prior to 1980. From 1980 on, we consider the 
double-Gaussian distribution as the most plausible model (Fig.~\ref{fig-bootstrap}). 

The double-log normal distribution of total US gross was previously reported in 2010 by Pan and Sinha for films released between 
1999 and 2008~\cite{Pan2010}. In a 2013 paper, Chakrabarti and Sinha demonstrated that such bimodality can arise in a stochastic 
model where theaters independently decide on which films to show~\cite{Chakrabarti2013}. We believe that the lower peak in the log 
total gross distribution includes big-budget films that flopped as well as ``independent'' films and ``art'' films that do not have 
the circulation afforded major studio releases. The appearance of the lower mode in the period 1980--1984 (Fig.~\ref{fig-bootstrap}) 
corresponds to a time of rapid growth in the size and number of movie theaters in the United States, including small independent 
film theaters. Between 1980 and 2000, the number of movie screens more than doubled as multiplexes replaced single-auditorium 
theaters~\cite{Noam2009}. In addition, during the early 1980s, art theaters began to arise in smaller urban and suburban areas, 
whereas prior to that period, art theaters were only located in a select few major cities~\cite{McLane2002}. This expansion of 
movie theaters enabled more low-budget films to be viewed by the general public and to have the opportunity for financial success.

\begin{figure}[tbp]
\centering
\includegraphics[width=1.0\textwidth]{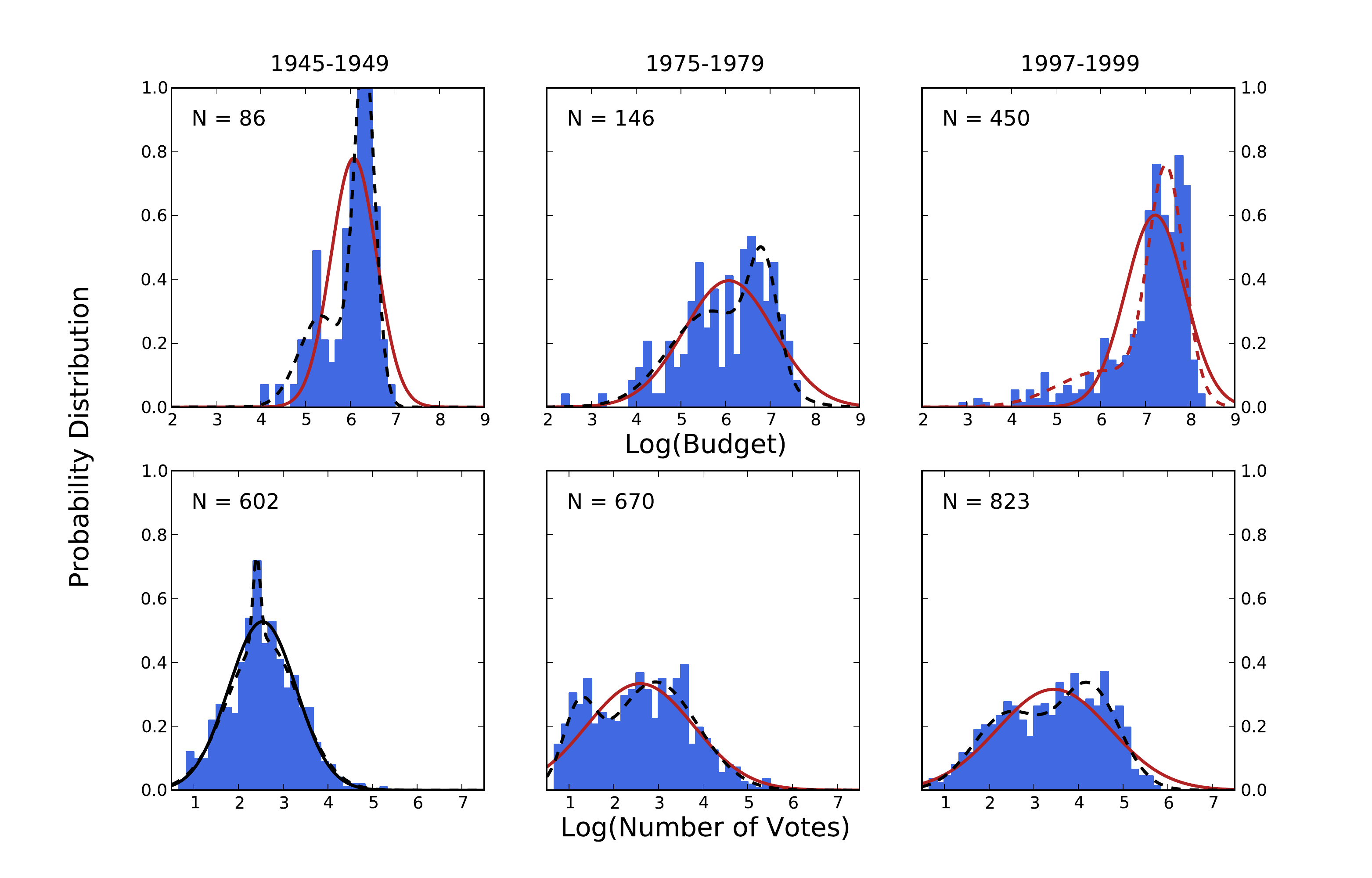}
\caption{
\textbf{Examples of distributions of log budget and log number of user votes.} 
Distribution of the log of budget (upper row) and the log of number of user votes (lower row) in different time windows. The 
lines represent the best fits of a single Gaussian distribution (solid line) and a mixture of two Gaussian distributions 
(dashed line) to the data. The color of the curve signifies whether we reject (red) or fail to reject (black) the distribution as 
a possible fit for the data. The number in the upper-left corner is the total number of data points in the sample.
}
\label{fig-hist-bv}
\end{figure}

For most of the considered time windows, the log of budget data exhibits a double-Gaussian distribution (Fig.~\ref{fig-hist-bv}).
This is the case for all time windows between 1935 and 1979 (Fig.~\ref{fig-bootstrap}). After 1979, we reject both the single- and 
double-Gaussian distributions as potential fits (Fig.~\ref{fig-bootstrap}). We suspect that the rejection of the unimodal and bimodal 
fits may be caused by the appearance of additional modes in the data beginning in 1980. This would place the emergence of new 
modes around the time of a reduction in cost of film production equipment, such as stereo sound recorders~\cite{Enticknap2005}. 
These new modes persist through the 1990s, as filmmaking switched from analog to digital.

The distributions for the log of total number of votes received behave in a similar fashion. The data is initially (1920--1964) best 
represented by a Gaussian distribution followed by a period (1965--2001) where a mixture of two Gaussian distributions works well. 
After 2000, neither of the two proposed models is a good fit (Fig.~\ref{fig-bootstrap}). The rejection of both proposed distributions 
after 2002 aligns with a marked rise in Internet usage and a large traffic increase for IMDb.

\section*{Discussion}

The number of votes received by a film is plausibly related to its quality and its financial characteristics. Thus, we next investigate 
a linear regression model for the log of number of votes.
\begin{equation}
  v(k) = a_{0} + a_{1}~\hat{b}(k) + a_{2}~\hat{g}_{T}(k) + a_{3}~r(k), \label{eq-lr}
\end{equation}
where $v(k)$ is the log number of user votes received by a film $k$, $\hat{b}(k)$ is the year-normalized log budget, $\hat{g}_{T}(k)$ 
is the year-normalized log total box office gross, and $r(k)$ is the average user rating. We normalize the monetary statistics by year 
in order to account for inflation, increases in ticket prices, and population growth. The procedure for calculating year-normalized 
log budget consists of subtracting a film's log budget value by the median of log budgets for all films released in the same calendar 
year,
\begin{equation}
  \hat{b}(k) = b(k) - \tilde{b}_{y(k)} \label{eq-norm}
\end{equation}
where $b(k)$ is the actual log budget of film $k$, $\tilde{b}_{y}$ is the median of log budgets for films released in year $y$, and 
$y(k)$ is the year of release. Normalized total gross is computed in the same fashion as in Eq.~\eqref{eq-norm}.

In order to properly estimate the parameter values for the model, we must account for the high prevalence of missing metadata. To do this, 
we use the Heckman correction method~\cite{Heckman1976,Heckman1979} to adjust for selection bias caused by the absence of financial or 
voting data in approximately two-thirds of films in the dataset. For this method, we utilize a linear probit selection model to 
estimate the probability that a film in the dataset reported the necessary information,
\begin{equation}
  \Pr(\text{data reported}) = d_{0} + d_{1}~y(k) + d_{2}~\left[i(k)\right]^{\frac{1}{3}} + d_{3}~\left[o(k)\right]^{\frac{1}{3}}, \label{eq-probit}
\end{equation}
where $i(k)$ is the in-degree of film $k$, and $o(k)$ is its out-degree. This conditional probability is then applied as a 
correction term to Eq.~\eqref{eq-lr}.

\begin{table}[tb]
\scriptsize
\centering
\begin{tabular}{l cc cc cc cc cc}
\textbf{Model} & \multicolumn{2}{c}{$r$ only} & \multicolumn{2}{c}{$\hat{g}_{T}$ only} & \multicolumn{2}{c}{$\hat{b}$ only} & \multicolumn{2}{c}{$\hat{b}$ and $r$} & \multicolumn{2}{c}{all} \\
\cmidrule(lr){2-3} \cmidrule(lr){4-5} \cmidrule(lr){6-7} \cmidrule(lr){8-9} \cmidrule{10-11}
& Coeff. & Std.Err. & Coeff. & Std.Err. & Coeff. & Std.Err. & Coeff. & Std.Err. & Coeff. & Std.Err. \\
\midrule
\textbf{Log of user votes} &&&&&&&&&&\\
~~Intercept & \textbf{3.16} & 0.05 & \textbf{4.67} & 0.02 & \textbf{4.85} & 0.03 & \textbf{4.19} & 0.04 & \textbf{3.80} & 0.04 \\
~~Log Norm. Budget $\hat{b}$ & --- & --- & --- & --- & \textbf{0.602} & 0.006 & \textbf{0.586} & 0.006 & \textbf{0.268} & 0.01 \\
~~Log Norm. Total Gross $\hat{g}_{T}$ & --- & --- & \textbf{0.354} & 0.006 & --- & --- & --- & --- & \textbf{0.191} & 0.008 \\
~~User Rating $r$ & \textbf{0.080} & 0.005 & --- & --- & --- & --- & \textbf{0.102} & 0.006 & \textbf{0.156} & 0.006 \\
&&&&&&&&&&\\
\textbf{Probit selection}&&&&&&&&&&\\
~~Intercept & 0.348 & 1.5 & \textbf{-80} & 2 & \textbf{-46} & 1 & \textbf{-46} & 1 & \textbf{-76} & 2 \\
~~Year $y$ & $\text{-}5 \!\!\times\!\! 10^{\text{-}5}$ & $8 \!\!\times\!\! 10^{\text{-}4}$ & \textbf{0.040} & $8 \!\!\times\!\! 10^{\text{-}4}$ & \textbf{0.023} & $6 \!\!\times\!\! 10^{\text{-}4}$ & \textbf{0.023} & $6 \!\!\times\!\! 10^{\text{-}4}$ & \textbf{0.037} & $9 \!\!\times\!\! 10^{\text{-}4}$ \\
~~Cube Root In-Deg. $i^{\frac{1}{3}}$ & \textbf{1.19} & 0.07 & \textbf{0.916} & 0.02 & \textbf{0.713} & 0.02 & \textbf{0.713} & 0.02 & \textbf{0.884} & 0.02 \\
~~Cube Root Out-Deg. $o^{\frac{1}{3}}$ & \textbf{0.915} & 0.06 & \textbf{0.354} & 0.02 & \textbf{0.250} & 0.02 & \textbf{0.250} & 0.02 & \textbf{0.317} & 0.02 \\
&&&&&&&&&&\\
~~Inverse Mills Ratio & \textbf{-6.47} & 0.2 & \textbf{-0.81} & 0.02 & \textbf{-1.16} & 0.02 & \textbf{-1.12} & 0.02 & \textbf{-0.61} & 0.01 \\
&&&&&&&&&&\\
Total Films & \multicolumn{2}{r}{15,425~~~~~~~~~} & \multicolumn{2}{r}{15,425~~~~~~~~~} & \multicolumn{2}{r}{15,425~~~~~~~~~} & \multicolumn{2}{r}{15,425~~~~~~~~~} & \multicolumn{2}{r}{15,425~~~~~~~~~} \\
Films w/ Observed Data & \multicolumn{2}{r}{14,577~~~~~~~~~} & \multicolumn{2}{r}{5,307~~~~~~~~~} & \multicolumn{2}{r}{5,331~~~~~~~~~} & \multicolumn{2}{r}{5,331~~~~~~~~~} & \multicolumn{2}{r}{3,430~~~~~~~~~} \\
Films w/ Censored Data & \multicolumn{2}{r}{848~~~~~~~~~} & \multicolumn{2}{r}{10,118~~~~~~~~~} & \multicolumn{2}{r}{10,094~~~~~~~~~} & \multicolumn{2}{r}{10,094~~~~~~~~~} & \multicolumn{2}{r}{11,995~~~~~~~~~} \\ 
&&&&&&&&&&\\
\textbf{Adjusted R-Squared \%} & \multicolumn{2}{c}{\textbf{26.02}} & \multicolumn{2}{c}{\textbf{63.74}} & \multicolumn{2}{c}{\textbf{73.97}} & \multicolumn{2}{c}{\textbf{75.50}} & \multicolumn{2}{c}{\textbf{72.83}} \\
\end{tabular}
\caption{
\textbf{Correlations among year-normalized financial data, user rating, and user votes.}
Results of correlation analyses comparing the log of number of user votes to the log of normalized budget, the log of normalized total 
gross, and the average user rating, using the Heckman correction method to account for missing data. Each column represents a different 
implementation of the linear model in Eq.~\eqref{eq-lr}. Due to limited space in the table, we do not display results for two of the 
possible models: The model using budget and total gross as independent variables, and the model using total gross and user rating as 
independent variables. We omit the budget and gross model because it is outperformed by the model using only budget (adjusted R-squared 
of 66.59\% vs. 73.97\%) as budget and gross are highly correlated. We omit the gross and rating model because it is outperformed by the 
model using budget and user rating (68.31\% vs. 75.50\%). Bolded coefficient values are significant with $p < 0.001$. 
These results suggest that budget alone explains much of the variation in number of votes. 
}
\label{tbl-votes-corr}
\end{table}

From the correction analysis, we find that both year-normalized financial statistics correlate strongly with total number of user votes 
(Tbl.~\ref{tbl-votes-corr}). The stronger correlation exists between the log of number of user votes and the log of normalized film budgets. 
When both business statistics are used in a linear model for number of votes, the correlation is not as strong as when the log of budget 
is used alone. The strongest correlation, however, exists for the use of log of normalized budget in conjunction with average user rating 
(Tbl.~\ref{tbl-votes-corr}). Interestingly, when we replace log of user votes with user rating as the dependent variable, we find no 
correlation with either financial quantity.

Our result suggests that the number of user votes is an indicator of a film's prominence. After all, a person is more likely to enter a 
rating for a film if they have viewed it, regardless of whether they found the film good or bad. However, prominence is not necessarily 
tied to box office success, as many films become notable for other reasons, such as major film award nominations. Films can also become 
notable for especially poor performances at the box office (e.g., 1995's \textit{Cutthroat Island}, which cost \$98 million to produce 
and only grossed \$10 million). In addition, the production budget for a film--also known as the `negative cost'--has been found to be 
strongly correlated with a film's advertising cost~\cite{Prag1994}. Hence it is understandable that total number of user votes correlates 
more strongly with budget than with box office gross, as the former is directly related to the amount spent on a film's promotion, which 
increases its prominence. Moreover, we find that film quality---in the form of average user rating---does not appreciably account for 
prominence when applied in conjunction with budget in the linear model. Therefore, budget is overwhelmingly the most relevant factor in 
determining a film's ultimate prominence. To make a film more notable, Hollywood does not need to spend more money on making it better; 
Hollywood just needs to spend more money.

\bibliography{imdb_citations_base}

\end{document}